\documentclass[letterpaper,twocolumn,10pt]{article}
\usepackage{usenix-2020-09}

\usepackage{booktabs}
\usepackage{pifont} 
\usepackage[most]{tcolorbox}
\usepackage{enumitem}
\usepackage{orcidlink}
\usepackage{tabularx}

\definecolor{cred}{HTML}{EAF2FF}     
\definecolor{pii}{HTML}{FFF2E6}      
\definecolor{netid}{HTML}{E6F4EA}    
\definecolor{cost}{HTML}{EFE9FF}     

\definecolor{selected}{HTML}{AAFAE7} 

\definecolor{toprow}{HTML}{FFF7CC} 

\renewcommand{\arraystretch}{1.12}
\usepackage[font=footnotesize, labelfont=bf]{caption}
\usepackage[table,dvipsnames]{xcolor}
\usepackage{balance}
\definecolor{lightred}{HTML}{FF6666}
\newcommand*\emptycirc[1][1ex]{\tikz\draw (0,0) circle (#1);} 
\newcommand*\halfcirc[1][1ex]{%
  \begin{tikzpicture}
  \draw[fill] (0,0)-- (90:#1) arc (90:270:#1) -- cycle ;
  \draw (0,0) circle (#1);
  \end{tikzpicture}}
\newcommand*\fullcirc[1][1ex]{\tikz\fill (0,0) circle (#1);} 

\begin{document}

\title{\Large \bf You Have Been LaTeXpOsEd: A Large-Scale Systematic Analysis of Information Leakage in Preprint Archives Using Large Language Models}

\author{
{\rm Richard A. Dubniczky \orcidlink{0009-0003-3951-1932}} \\
E\"otv\"os Lor\'and University \\
dubniczky@inf.elte.hu
\and
{\rm Bertalan Borsos \orcidlink{0009-0000-8718-8285}} \\
E\"otv\"os Lor\'and University \\
fshntj@inf.elte.hu
\and
{\rm Tamas Bisztray \orcidlink{0000-0003-2626-3434}} \\
University of Oslo, HUN-REN Sztaki\\
bisztray.tamas.gyorgy@sztaki.hu
\and
{\rm Norbert Tihanyi \orcidlink{0000-0002-9002-5935}} \\
Technology Innovation Institute \\
norbert.tihanyi@tii.ae
}

\maketitle
\begin{abstract}
In this work, we present the first large-scale security audit of the arXiv preprint repository, analyzing over 1.2 TB of data from 100,000 arXiv submissions to report on systemic sensitive information leakage. When authors upload submissions, they publish not only a PDF but also auxiliary code, images, and LaTeX source files containing embedded comments. In the absence of sanitization, these files often disclose sensitive information that adversaries can harvest using open-source intelligence. Operating under a strict ethical framework of passive verification, we introduce LaTeXpOsEd, a pipeline that integrates pattern matching, logical filtering, and large language models (LLMs) to detect context-dependent secrets within LaTeX comments and unreferenced auxiliary files. To evaluate the secret-detection capability of LLMs, we introduce LLMSec-DB, a benchmark on which we tested 25 state-of-the-art models. Analyzing publicly available arXiv submissions, we uncover thousands of PII exposures, hundreds of instances of exposed credentials, private Google Drive links, API keys, and various semantic leaks, including internal disputes and confidential peer reviews. We show that this large-scale extraction of sensitive information is economically viable for low-resource adversaries leveraging open-weight models and constitutes a serious security and reputational threat to individuals and institutions. We urge the research community and repository operators to take immediate action to close these hidden security gaps. To support open science and in accordance with responsible disclosure standards, we have published our toolset and benchmarks\ on GitHub\footnote{\url{https://github.com/LaTeXpOsEd}}  and Zernodo\footnote{\url{https://doi.org/10.5281/zenodo.20035860}}\footnote{\url{https://doi.org/10.5281/zenodo.20058989}}.
\end{abstract}

\section{Introduction}

Scientific publishing has expanded rapidly, increasing from roughly 1 million papers per year in the early 2000s to nearly 2.9 million by 2020~\cite{nsf2021,10.1162/qss_a_00327}. In 2025, growing pressure to communicate results quickly has made knowledge sharing faster than ever, but the traditional peer-review cycle---often taking months or years---can delay impact and render work outdated, especially in fast-moving fields such as \textit{artificial intelligence (AI)}, biotechnology, quantum computing, and genomics.
To reduce this lag, preprint repositories such as arXiv have become central to scientific communication by allowing researchers to share results immediately, well before journal publication. Preprints are freely accessible, support open science, and provide a timestamped DOI record that establishes authorship and priority.
Unlike most journals, which typically publish only a final PDF, arXiv requires submission of the underlying source files (e.g., LaTeX and build artifacts), improving transparency and supporting long-term reproducibility~\cite{nsf2021}. This architectural choice enables a new vector for \textit{Open-Source Intelligence (OSINT)}~\cite{OSINT}. Source files often contain hidden artifacts like commented-out credentials, API keys, internal comments, and metadata. Treating these files as ephemeral build artifacts is a misconception, both from a legal and structural security standpoint. By accepting the repository's submission agreement, authors grant an ``\textit{irrevocable}'' and ``\textit{perpetual}'' license to distribute these materials as integral components of ``The Work''~\cite{arxiv_submission_agreement, arxiv_terms}. The platform's Privacy Policy explicitly shifts the burden of sanitization to the submitter, stating that users ``\textit{must ensure that [...] inclusion of the article in arXiv does not violate the rights of any person or entity, including privacy rights}''~\cite{arxiv_privacy}. The provision of bulk data access via Amazon S3 enables large-scale computational analysis of this data, negating any practical obscurity~\cite{arxiv_s3}.

While simple pattern matching (regex) can identify structured data like IP addresses, it fails to detect context-dependent sensitive information, such as author disputes, semantic leaks, or credentials hidden within comments. Detecting such exposures at scale has not been possible historically. The emergence of \textit{Large Language Models (LLMs)}, which consistently achieve state-of-the-art performance across a wide range of \textit{Natural Language Processing (NLP)} tasks, makes them exceptionally effective for this kind of analysis. 

\noindent LLMs' ability to understand context, extract meaning, and interpret unstructured text enables them to uncover hidden or unintended disclosures within complex scientific documents.

\begin{figure*}[t]
\centering
\includegraphics[width=\textwidth]{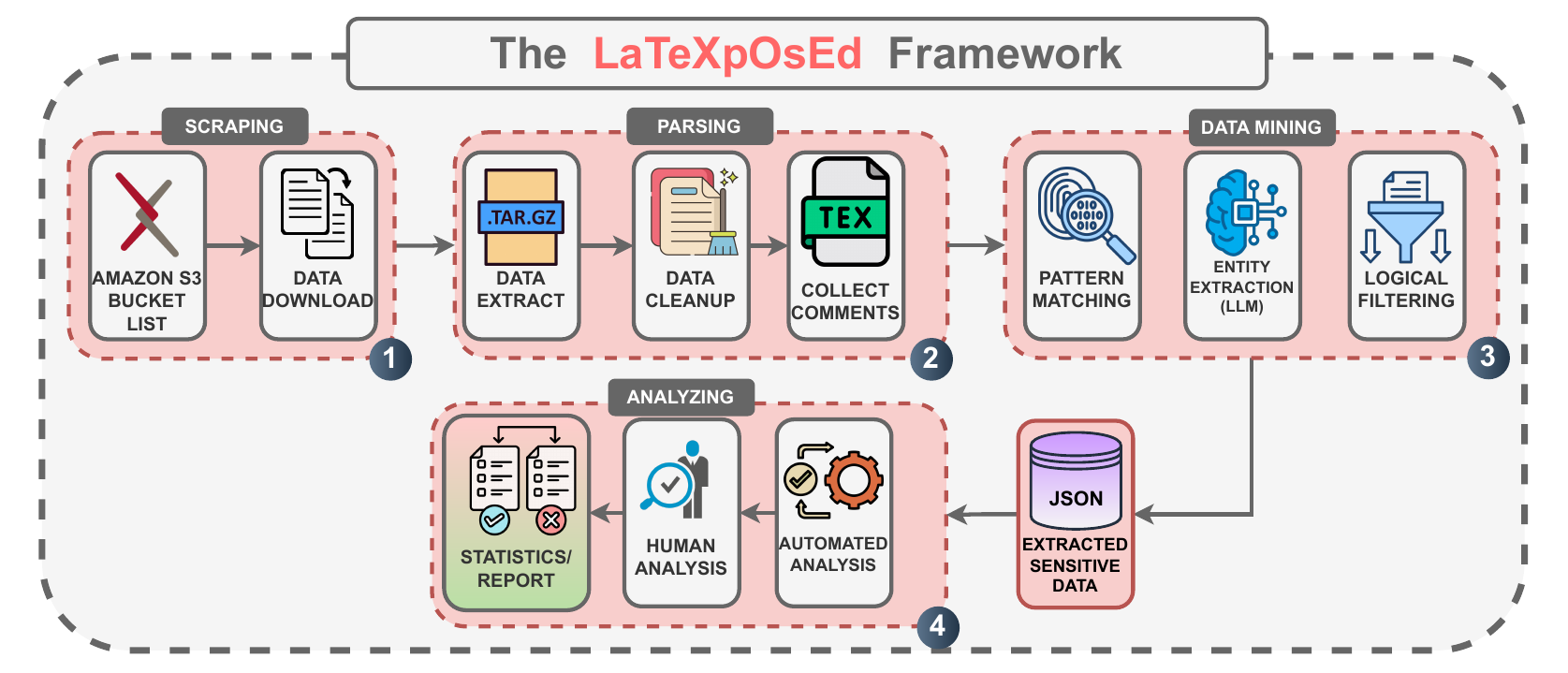}
\caption{The LaTeXpOsEd framework: a four-step process for scraping, parsing, mining, and analyzing documents.}
\label{fig:Framework}
\end{figure*}
To quantify the extent of sensitive data exposure in the scholarly research artifacts, and to assess whether LLMs are suitable for the extraction of context-dependent secrets, our analysis is guided by two primary research questions:

\begin{tcolorbox}[colback=gray!10, colframe=black, arc=6pt, boxrule=0.7pt, left=2mm, right=2mm, top=1mm, bottom=1mm, rounded corners]
\begin{enumerate}
    \item \textbf{RQ1}: What kinds of sensitive information are exposed in preprint source files and comments, and how often do these disclosures occur?
    \item \textbf{RQ2}: How effective are LLMs at detecting contextually hidden sensitive data within preprint source materials?
\end{enumerate}
\end{tcolorbox}

To answer these questions, we present the \texttt{LaTeXpOsEd} framework, which consists of four phases, as shown in Figure~\ref{fig:Framework}. Our main contributions are as follows:

\begin{itemize}
    \item \textbf{The \texttt{LaTeXpOsEd} Framework}: We propose a new methodology that integrates entity extraction, pattern matching, and the contextual understanding of LLMs to systematically process LaTeX source files. This approach uncovers context-dependent secrets—such as credentials in comments, author disputes, and semantic leaks—missed by automated auditing.

    \item \textbf{Large-Scale Preprint Analysis}: We conduct the first comprehensive security audit of preprint source files, examining over $1.2$ TB of data across $100\,000$ arXiv submissions. We systematically analyze comments and unreferenced auxiliary materials.

    \item \textbf{\texttt{LLMSec-DB} Benchmark}: To evaluate the secret-detection capabilities of LLMs, we introduce a new benchmark based on realistic synthetic samples. We evaluate 25 state-of-the-art LLMs, and this evaluation serves as the basis for selecting the optimal model for auditing the arXiv dataset using a weighted analysis of cost efficiency, accuracy, and open-source availability.

    \item \textbf{Automated Data Analysis with LLMs}: We uncovered thousands of instances of \textit{Personally Identifiable Information} (PII), hundreds of critical credential leaks, API keys, private Google Drive access links, and sensitive conversations unintentionally exposed in source files.
\end{itemize}

To support open science, we have made the \texttt{LLMSec-DB} dataset and all scripts used in this research publicly available on the project website: \url{https://github.com/LaTeXpOsEd}.

This paper is structured as follows: Section~\ref{sec:ethics} investigates the ethical and legal background, Section~\ref{sec:related} overviews the related work, Section~\ref{sec:methodology} presents the methodology, while Section~\ref{sec:discussion} discusses the results of the benchmarking and scraping activities. Section~\ref{sec:limit} outlines limitations, and finally, Section~\ref{sec:conclusion} summarizes our results.

\section{Ethical Considerations}
\label{sec:ethics}
Ethical considerations and responsible disclosure are central to this research. Our methodology is grounded in a \textit{Compliance Stack} that harmonizes international regulations, principles of computing ethics, and arXiv’s Terms of Use. Under arXiv’s \textit{Submission Agreement},  authors grant the platform a ``non-exclusive, perpetual, irrevocable'' license to distribute submitted works. Furthermore, the \textit{Privacy Policy} places the responsibility on submitters, explicitly warning them to remove any content they do not wish to make public. From a regulatory perspective both the \textit{US Common Rule (Exemption 4)} and \textit{GDPR (Article 9(2)(e))} provide legal basis for processing data that has been ``\textit{manifestly made public}'' by the subject. To operationalize these principles and strictly rule out the possibility of causing harm, we adhered to the following guidelines:
\begin{itemize}

\item \textbf{Data Acquisition:} All downloads from arXiv servers were performed exclusively through arXiv’s official Amazon S3 bulk data access service~\cite{arxiv_s3}—the platform’s recommended mechanism for large-scale data acquisition—thereby avoiding any practices resembling abusive scraping or denial-of-service activities.

\item \textbf{Passive Verification:} In accordance with the \textit{Menlo Report}~\cite{menlo_report}, a cybersecurity ethics framework developed with support from the \textit{U.S. Department of Homeland Security} (DHS), we adopted a strict passive-verification framework. We did not attempt to validate leaked credentials (usernames, passwords, API keys) against live systems. For exposed URLs, we limited our analysis to reachability checks without scraping or sharing any private content or the links themselves.

\item \textbf{Data Privacy and Anonymization:} No sensitive data or PII is disclosed. Findings are anonymized, aggregated and later deleted, to prevent re-identification. Our goal is to highlight systemic infrastructure risks, not to expose specific individuals. Additionally, downloaded papers were randomized both in selection and in order to prevent the linkage of results to papers. On this benchmark the generated result shoudl match toally with th epresdneted resulty in the apper (

\item \textbf{Responsible Disclosure:} We formally notified the arXiv administration of the systemic nature of these disclosures on December 12, 2025, emphasizing the associated security risks and potential reputational harm. ArXiv acknowledged our notification on December 16, 2025 and internally shared the report with the relevant teams for further discussion and assessment. 
\item \textbf{Authors Notification:} Given the scale of the study, individual notification of all affected authors and organizations is infeasible. For critical findings, such as leaked credentials or publicly shared private links (e.g., Dropbox or Google Drive), we notified corresponding authors whenever identification and contact were possible.
\end{itemize}

\section{Related Literature}
\label{sec:related}

\subsection{The Landscape of Source Mining}

While arXiv has long been used for bibliometric analysis, prior work has largely relied on metadata or PDF extraction (e.g., unarXive)~\cite{saier2020unarxive}. More recently, researchers have begun mining the underlying LaTeX source, which preserves behavioral signals lost in the compiled PDF. Pei et al.\ analyzed 1.6 million LaTeX submissions to infer collaboration structure from author-specific macros and commented-out sections~\cite{pei2025hidden}. This demonstrates the feasibility of source-level analysis at scale, but their focus is sociological rather than security-related. We adopt the same source-first perspective and apply it to the study of information leakage. Table~\ref{tab:repositories} underscores why arXiv is central in this setting. Many repositories act as PDF-only archives---implicitly removing source-level artifacts such as comments and auxiliary files---or make source uploads optional. By contrast, arXiv requires source submission for LaTeX-based papers. With more than 2.8 million entries, it is one of the few platforms that combine mandatory source availability with a corpus of millions of papers, making it uniquely suited for large-scale leakage analysis.

Circling back to user responsibility, the presence of sanitization tools such as arxiv-latex-cleaner~\cite{google_research_arxiv_latex_cleaner} (Google Research) and latexpand~\cite{latexpand_gitlab} indicates that if the community would understand the risks, mitigation would be rather straightforward. These tools can remove comments and auxiliary files prior to submission. When authors do not use available sanitization tools, despite explicit warnings by arXiv to ``
\textit{Tidy your submission}'', any resulting exposure is best understood as a preventable consequence of that choice.

\subsection{Information Leakage in Open Repositories}
Metadata and file-level side channels are not unique to archiving tools. Researchers demonstrated that different versions of social media images often retain EXIF data (GPS, timestamps) that act as side channels for context leakage~\cite{10.1007/s11280-025-01335-1,Soni2025}.

Users frequently misunderstand that archiving creates a permanent, searchable record where accidental inclusions become irrevocable security liabilities. To date, arXiv source uploads have received limited attention with respect to potential privacy and security relevant information leakage. On the contrary, GitHub in particular, has been identified as a primary vector for credential leakage~\cite{10063545}, driving a robust body of forensic literature. Meli et al. performed a large-scale longitudinal scan of billions of files on GitHub and found secret leakage to be pervasive---impacting over 100{,}000 repositories---with thousands of new, unique secrets exposed each day, many of which remain publicly accessible for weeks or longer~\cite{Meli2019}.
\begin{table}[h!]
\footnotesize
    \centering
    \caption{Comparison of source file availability across preprint repositories.}
    \rowcolors{2}{gray!10}{white}
    \renewcommand{\arraystretch}{1.2} 
    \setlength{\tabcolsep}{8pt}      
    \begin{tabular}{p{1.5cm}|p{2.7cm}|c|r}
        \rowcolor{gray!50}
        \textbf{Archive} & \textbf{Owner} & \textbf{Source} & \textbf{Entries} \\
        \toprule
        \textbf{arxiv.org} & Cornell University      & \fullcirc  & $>2.8$M \\
        ssrn.com           & Elsevier                & \emptycirc & $>1.7$M \\
        hal.science        & CCSD                    & \halfcirc  & $>1.5$M \\
        zenodo.org         & CERN                    & \halfcirc  & $>100$k \\
        preprints.org      & MDPI                    & \emptycirc & $>100$k \\
        osf.io             & Center for Open Science & \halfcirc  & $>25$k \\
        figshare.com       & Digital Science         & \halfcirc  & $>5.4$k \\
        techrxiv.org       & IEEE                    & \halfcirc  & $>1$k \\
        \bottomrule
    \end{tabular}
    \smallskip
    \scriptsize
    \textbf{Legend for Source File Availability:} \fullcirc\ = Mandatory for LaTeX submission \\ \halfcirc\ = Optional source upload \emptycirc\ = PDF-only, source not provided.
    \label{tab:repositories}
\end{table}

\subsection{Evolution of Secret Detection Techniques}
Detecting hardcoded secrets such as credentials, API keys, and PII in source code has been extensively studied, with a wide range of approaches proposed in the literature, ranging from traditional regular-expression–based techniques to more advanced methods incorporating entropy analysis and data-flow analysis.

Yayin et al. introduced CredMiner~\cite{CredMiner}, a framework for identifying developer credentials unsafely embedded in Android applications. By combining data-flow analysis with selective execution, CredMiner can recover even obfuscated credentials at scale. Feng et al. introduced PassFinder, a deep learning approach that uncovered over $60{,}000$ password leaks in GitHub repositories, demonstrating the scale of credential exposure in software repositories~\cite{9794113}. Basak et al. evaluated popular tools on GitHub such as TruffleHog and GitLeaks, finding that regex-based approaches suffer from high false-positive rates and struggle to capture contextual information~\cite{10304853}. Their manual analysis further showed that false positives largely result from overly generic regular expressions and ineffective entropy-based heuristics, while false negatives stem from faulty or incomplete patterns, skipped file types, and insufficient rulesets.

\subsubsection{LLM-based secret detection}
Prior research has shown that traditional regex-based methods often produce a high number of false positives and suffer from significant false negatives, particularly when dealing with non-standard or context-dependent information leaks. To address these limitations, researchers have increasingly explored the use of LLMs, which are capable of understanding contextual information and semantic meaning, enabling more accurate discrimination between genuine secret leakages and benign code artifacts.

Yang et al. explored the utility of LLMs for privacy protection by introducing the \textit{CRSRF prompting framework} to detect PII in archival records using ChatGLM2-6B~\cite{10386949}. Their study demonstrated that unsupervised LLMs could achieve high precision in identifying traditional PII—such as names, addresses, and ID numbers—while providing human-interpretable rationales for their decisions. However, their evaluation relied on a small, manually curated dataset of 300 samples and focused on demographic data, without assessing the capability of LLMs to detect technical secrets.

Alecci et al. proposed \textit{SecretLoc}, using LLMs to detect secrets in Android apps without predefined patterns and successfully identifying novel key types (e.g., OpenAI API keys) that evaded rule-based scanners~\cite{alecci2025evaluating}. Similarly, Rahman et al. proposed a hybrid regex+LLM pipeline for source code, achieving an F1 score of $\approx 0.985$ and substantially reducing false positives compared to regex-only baselines~\cite{rahman2025secretbreachdetectionsource}.

Recently, Biringa and Kul presented a comprehensive study on detecting hard-coded credentials in software repositories using LLMs~\cite{biringa}. Their approach leverages transformer-based contextual embeddings extracted from pre-trained LLMs and feeds them into a deep learning classifier, outperforming state-of-the-art methods. The study demonstrates that LLMs significantly reduce false positives by effectively capturing contextual and semantic dependencies that embedding-only and rule-based approaches fail to model.

Key contributions spanning traditional secret detection techniques to large-scale LLM-based methods are summarized in Table~\ref{tab:relatedwork}.
While LLM-based approaches have achieved exceptional results in secret detection within software repositories, they do not address whether information leakage occurs in arXiv source files, which types of secrets are most prevalent in this context, or which LLMs (if any), can reliably extract these specific secrets.

\begin{table}[htb]
\footnotesize
\centering
\caption{Summary of related work on repository mining, secret detection, and LLM-based secret analysis}
\rowcolors{2}{gray!10}{white}
\renewcommand{\arraystretch}{1.3} 
\setlength{\tabcolsep}{6pt}
\begin{tabular}{p{0.8cm}|p{2.2cm}|p{4.5cm}}
\rowcolor{gray!60}
\textbf{Year} & \textbf{Study} & \textbf{Contribution} \\
\toprule
2015 & Yayin et al. & Introduced CredMiner, a framework for detecting embedded developer credentials in Android applications.~\cite{CredMiner}. \\
\hline
2020 & Saier et al. & Constructed unarXive, a large-scale dataset of 1M+ arXiv papers with citation contexts~\cite{saier2020unarxive}. \\
\hline
2022 & Feng et al.  & Detected hardcoded passwords in 60k+ GitHub repositories using deep learning~\cite{9794113}. \\
\hline
2023 & Basak et al. & Demonstrated high false-positive rates and limited recall in standard regex-based secret-detection tools~\cite{10304853}. \\
\hline
2023 & Yang et al.  & Applied LLMs to identify PII in massive archival text datasets with human-interpretable rationales~\cite{10386949}. \\
\hline
2025 & Pei et al.   & Extracted hidden author contributions from 1.6M arXiv LaTeX source files to study collaboration norms~\cite{pei2025hidden}. \\
\hline
2025 & Alecci et al.& Proposed \textit{SecretLoc}, an LLM-based method to detect obfuscated secrets in Android applications~\cite{alecci2025evaluating}. \\
\hline

2025 & Rahman et al.& Proposed a hybrid regex+LLM approach for secret detection in source code~\cite{rahman2025secretbreachdetectionsource}. \\
\hline
2025 & Biringa et al. & Leveraged transformer-based LLM embeddings combined with a deep learning classifier to detect hard-coded credentials in software repositories, achieving improved precision and reduced false positives over state-of-the-art methods.~\cite{biringa} \\

\hline
\rowcolor{red!10}
\textbf{2025} & \textbf{Our Work} & To the best of our knowledge, the first systematic security \textbf{audit of arXiv} source files, and the first \textbf{standardised benchmark} to facilitate a comparative analysis of LLMs to extract PII, credentials, and context dependent comments. \\
\bottomrule
\end{tabular}
\label{tab:relatedwork}
\end{table}

\section{Methodology}
\label{sec:methodology}

Our methodology is divided into four main phases, as illustrated in Figure~\ref{fig:Framework}. Scraping (collecting the data), Parsing (structuring and cleaning the source files), Data Mining (applying pattern matching and LLM-based extraction), and Analyzing (categorizing and interpreting the findings).

\subsection{Phase 1 - Scraping}

To download the LaTeX source files, we utilised arXiv's Amazon S3 bulk data service.
By accessing \texttt{tar.gz} archives (containing $\approx 125$ papers each) and utilizing the provided XML index, we identified $808$ specific archives required for our dataset. 
This method allowed us to retrieve the $100{,}000$ papers, totaling $400$~GB of compressed data ($1.2$~TB uncompressed). Because the arXiv S3 bucket is configured as a Requester Pays bucket, data transfer and request charges were billed to our AWS account, resulting in a cost of \$37 in research expense.

\subsection{Phase 2 - Parsing}

\begin{figure}[b]
  \centering
  \includegraphics[width=0.8\columnwidth,page=1]{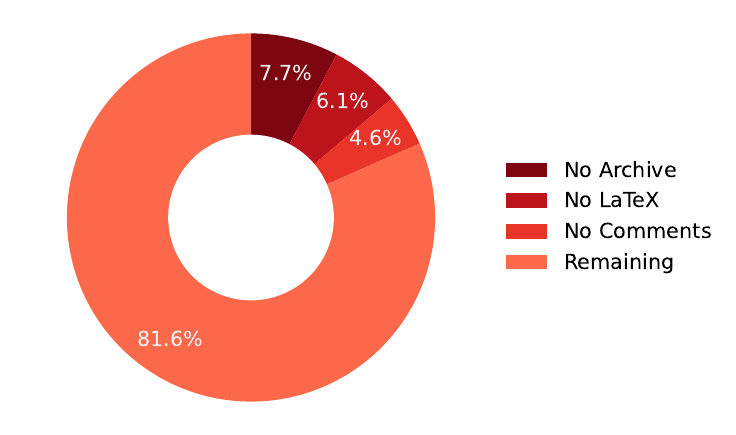}
  \caption{Ratio of papers without usable comments in LaTeX source files.}
  \label{fig:comment_ratio}
\end{figure}

In the parsing stage, we decompressed the \texttt{tar.gz} archives and extracted relevant content into machine-readable JSON files to support downstream analysis. Overall, approximately 18\% of the analyzed papers either lack usable LaTeX comments for subsequent processing or provide no source archive. Figure~\ref{fig:comment_ratio} shows a detailed breakdown of the subcategories.
Approximately 7.7\% of downloaded submissions did not include a source archive and were available only as PDFs. Among the remaining submissions, 6.1\% contained source archives but no LaTeX files (e.g., non-LaTeX formats or incomplete sources). A further 4.6\% included LaTeX code but contained no comments. This final category is particularly notable, as the absence of comments is consistent with systematic removal, potentially reflecting manual editing practices or automated sanitization pipelines driven by institutional policies governing document preparation and public release.

To better understand the origin of these sanitized submissions (approximately 4{,}600 papers), we applied named entity recognition (NER)~\cite{keraghel2024recentadvancesnamedentity} to extract affiliation metadata, focusing on countries and organizations. The most frequent submitting countries were the United States (25\%), China (15\%), and Germany (7\%). These proportions closely match arXiv's reported submission distribution~\cite{arxiv_blog_12k_may_2018}, suggesting that no single country dominates systematic sanitization. At the organizational level, however, large industrial research institutions were disproportionately represented among submissions with evidence of comment removal, including Intel, Google, AWS, NVIDIA, IBM, Microsoft, and Apple.

Across the remaining 92{,}000 papers, we collected over 343{,}000 LaTeX source files containing more than 12 million comments. We then applied a multi-stage cleaning pipeline that removes empty entries, boilerplate, purely syntactic LaTeX commands, decorative separators, and redundant fragments. We also normalized whitespace and discarded comments appearing more than 10 times across the dataset, as these primarily consisted of common control sequences, formatting artifacts, or single words unlikely to contain sensitive content. After cleaning, the dataset was reduced to approximately 6.6 million comments of varying length, corresponding to just under 55\% of the original set. Comments are counted either as \textit{\textbackslash begin\{comment\}} blocks or as lines that start with the \texttt{\%} symbol.

This reduction is essential for scalable LLM-based analysis. Under the \texttt{CL100k\_base} encoding, the raw comment corpus contains approximately 2.4B tokens, whereas the cleaned set contains fewer than 275M tokens (an 89\% reduction). For context, processing the unfiltered corpus with an LLM such as GPT-5-mini would exceed \$3{,}000 at an input token rate of \$1.25 per million tokens (excluding output tokens). By contrast, the cleaned dataset reduces input-side cost to approximately \$344 and substantially lowers inference time.

For each paper, we exported all textual content into a dedicated dump file for systematic pattern matching. In parallel, we reconstructed each paper's logical structure to identify auxiliary files that are present in the source bundle but never referenced in the final manuscript. We note, that files included in the submission package but not referenced or used in the final compiled PDF are more likely to contain sensitive information. Because the contents of these files are invisible to readers and reviewers, authors may have a false sense of security and incorrectly assume that no information leakage has occurred.

To improve processing efficiency, we removed image files (e.g., JPEG, PNG) from the working set. Before deletion, however, we performed EXIF metadata extraction~\cite{EXIF} to detect embedded sensitive information. While many modern platforms strip EXIF data for privacy reasons, arXiv does not. These files therefore provide a valuable signal: we identified nearly 1{,}200 images containing sensitive metadata. The extracted fields vary widely. Even device and software identifiers (e.g., camera model, application versions) can be privacy-relevant.  Section~\ref{sec:discussion} provides a discussion about these findings.

\subsection{Phase 3 - Data Mining}

In this step we use pattern-matching techniques, LLMs, and logical narrowing techniques to find sensitive information. As illustrated in Figure~\ref{fig:Framework}, this stage involves three main processes: pattern matching, LLM-based extraction, and logical filtering.

\subsubsection{Pattern Matching} At this stage, pattern-matching techniques and regular expressions were applied to detect IP addresses, URLs, email addresses, banking information (such as IBANs), and other easily identifiable secrets in comments. The findings were then filtered and categorized manually and with LLMs. The remaining potentially sensitive links were separated into categories such as signed URLs, IP addresses, access tokens, and more. We utilized Trufflehog\footnote{\url{https://github.com/trufflesecurity/trufflehog}}, a secret detection tool originally designed for Git repositories, to scan for API keys, user credentials, tokens, along with $1,750$ secret patterns from the \textit{Secret Patterns Database Project}~\cite{secrets_patterns_db}.

We excluded all information from the final results that clearly indicated it was a temporary or illustrative finding. Examples include loopback IP addresses, URLs, and email addresses using the \textit{example.com} domain, and common tokens used for demonstration purposes. Running over $150$ million regex searches on all extracted comments took approximately $66$ minutes on a MacBook Pro with an M4 chip.

\subsubsection{Entity Extraction Benchmarking}
\label{sec:list}
LLMs were used to detect categories of sensitive information that cannot be reliably captured through pattern-based techniques. This includes conversational content such as internal disagreements between co-authors, critical remarks about the validity or rigor of methods and results, and partial or full references to peer-review reports and corresponding rebuttals.  

LLMs are also essential for detecting non-standard credentials that do not follow recognizable formats or regular expressions, for example statements such as “For the GitHub repository, the user is PROJECT-765 and the corresponding credential is Leaky5678.”  Identifying such disclosures is challenging without LLMs because it requires contextual interpretation and, in practice, multilingual understanding.

To evaluate LLM effectiveness for this task, we constructed a labeled dataset of 300 \emph{synthetic} comment snippets: 200 containing sensitive content and 100 benign. To ensure realistic coverage of secret types, we first sampled candidate snippets from a small set of arXiv sources to identify common disclosure patterns, and then recreated sanitized, synthetic versions that preserve structure and intent while removing any real secrets or identifying details. To emulate realistic operating conditions, we injected these synthetic snippets into the comment streams extracted from 1{,}000 PDF files, reflecting the degraded performance observed when models must process longer inputs~\cite{dubniczky2025castle}. We refer to this dataset as \texttt{LLMSec-DB}, and it is available for download from the project website for further use. Synthetic content was categorized into the following six classes:

\begin{itemize}
    \item \textbf{PII}: Personally identifiable information such as names, email addresses, phone numbers, or physical addresses. This also includes hidden author names or other data that could reveal an individual’s identity.  
    \item \textbf{CRED}: Real authentication material or payment data. Examples include passwords or passphrases, API keys and tokens, bearer tokens, client secrets, private keys, database connection strings with embedded credentials, and payment card information (PAN, expiry, CVV/CVC/CID).  
    \item \textbf{NETID}: System or network identifiers used to label accounts or machines, but not secrets themselves. Examples include usernames, user IDs, IP addresses, hostnames, workstation IDs, MAC addresses, and port numbers.  
    \item \textbf{PEER}: Content directly related to the formal peer review workflow, such as reviewer comments, meta-reviews, rebuttals, responses to reviewers, or camera-ready change requests. It also covers internal author discussions and strategy notes about how to respond to reviews, even if critical of reviewers.  
    \item \textbf{CONF}: Explicit disagreements or disputes among co-authors, for example, regarding methodology, content, tone, style, or the direction of the paper. This category does not include disagreements with reviewers.  
    \item \textbf{OTHER}: Miscellaneous sensitive content not covered by the above categories.
\end{itemize}

We benchmarked 25 LLMs on this dataset to identify a practical trade-off between detection quality and computational cost. Because some snippets contain multiple types of sensitive information, we report both \textit{Exact-Match Accuracy (EMA)}, which requires all labels to be correct, and \textit{At-Least-One Accuracy (ONE)}, which credits a prediction if it correctly identifies at least one relevant label. For our use case, ONE is the more operational metric, since any positive match triggers manual review. Consider the following log snippet:
\textit{``Windows EventID 4672, ACCOUNT: sec\_admin, PASSWORD: Pass!4682, MACHINE\_ID: WEB\_DEV-001''}.
If a model identifies only one class correctly---for example, network identifiers but not credentials---the snippet still warrants investigation and therefore counts as a valid finding in our pipeline.

\begin{table*}[t]
\centering
\caption{Performance of Tested LLMs on the LLMSec-DB Dataset (Sorted by Exact-Match Accuracy, EMA)}
\rowcolors{3}{gray!10}{gray!10} 
\begin{tabular}{lccc|ccc|ccc|ccc|cl}
\toprule

\multicolumn{4}{c}{} & \multicolumn{3}{c}{\textbf{CRED}} & \multicolumn{3}{c}{\textbf{PII}} & \multicolumn{3}{c}{\textbf{NETID}} & \multicolumn{2}{c}{\textbf{Costs}} \\
\cmidrule(lr){5-7} \cmidrule(lr){8-10} \cmidrule(lr){11-13} \cmidrule(lr){14-15}
\rowcolor{white}
\textbf{Model} & \textbf{EMA} & \textbf{ONE} & \textbf{O} & \textbf{TP} & \textbf{FP} & \textbf{FN} & \textbf{TP} & \textbf{FP} & \textbf{FN} & \textbf{TP} & \textbf{FP} & \textbf{FN} & \textbf{MT} & \textbf{ALL} \\
\midrule
GPT-5                       & 82.7\% & 91.7\% & \ding{55} & 50 &  3 & 0 & 42 & 16 & 2 & 38 &  9 & 4  & \$1.25 & \$840 \\
Claude Sonnet 4             & 81.0\% & 87.7\% & \ding{55} & 50 &  0 & 0 & 44 & 10 & 0 & 39 &  8 & 3 & \$3.00 & \$2016 \\
Claude 3.7 Sonnet           & 80.3\% & 92.0\% & \ding{55} & 50 &  0 & 0 & 43 & 21 & 1 & 37 & 12 & 4 & \$3.00 & \$2016\\
\rowcolor{red!20}
Qwen-2.5 72B                & 80.0\% & 91.0\% & \ding{52} & 50 &  2 & 0 & 44 & 11 & 0 & 38 & 11 & 3 & \$0.07 & \$47 \\
Llama-3.1 405B              & 80.0\% & 87.7\% & \ding{52} & 49 &  2 & 1 & 44 & 11 & 0 & 36 & 10 & 5 & \$0.80 & \$538 \\
Llama-3.3 70B              & 76.0\% & 86.3\% & \ding{52} & 49 &  5 & 1 & 44 &  7 & 0 & 35 & 12 & 6 & \$0.04 & \$27 \\
Gemini 2.5 Flash Lite       & 77.0\% & 86.0\% & \ding{55} & 45 &  1 & 5 & 44 &  8 & 0 & 33 & 10 & 8 & \$0.10 & \$67 \\
GPT-5 Nano                  & 76.7\% & 85.0\% & \ding{55} & 50 &  6 & 0 & 44 & 17 & 0 & 34 & 13 & 7 & \$0.05 & \$34 \\
Gemini 2.5 Pro              & 75.0\% & 93.3\% & \ding{55} & 50 &  0 & 0 & 42 & 24 & 2 & 32 & 14 & 8 & \$1.25 & \$840 \\
Deepseek R1                 & 75.0\% & 86.7\% & \ding{52} & 48 &  5 & 2 & 43 & 17 & 1 & 31 & 16 & 9 & \$0.40 & \$269 \\
GPT-5-mini                  & 71.3\% & 84.3\% & \ding{55} & 50 & 11 & 0 & 44 & 24 & 0 & 30 & 17 & 10 & \$0.25 & \$168 \\
NVIDIA Nemotron 9B          & 70.3\% & 80.3\% & \ding{52} & 50 & 15 & 0 & 44 & 19 & 0 & 29 & 18 & 11 & \$0.25 & \$168 \\
GPT-4o mini                 & 67.7\% & 86.3\% & \ding{55} & 50 & 13 & 0 & 43 &  9 & 1 & 28 & 17 & 13 & \$0.15 & \$101 \\
Grok4                       & 67.0\% & 81.7\% & \ding{55} & 49 &  0 & 1 & 42 & 20 & 2 & 27 & 20 & 14 & \$3.00 & \$2016 \\
Qwen3 235B-A22b             & 78.7\% & 86.0\% & \ding{52} & 48 &  3 & 2 & 42 &  2 & 2 & 33 &  8 & 8 & \$0.30 & \$202 \\
Gemma 3 27B                 & 65.7\% & 82.0\% & \ding{52} & 50 & 18 & 0 & 44 & 25 & 0 & 27 & 19 & 15 & \$0.07 & \$47 \\
Gemini Flash-1.5 8B            & 65.0\% & 74.3\% & \ding{55} & 48 & 10 & 2 & 43 &  5 & 1 & 25 & 20 & 17 & \$0.30 & \$202 \\
Mixtral 8x22b               & 63.3\% & 77.3\% & \ding{52} & 43 &  3 & 7 & 44 & 25 & 0 & 22 & 24 & 20 & \$0.90 & \$605 \\
Mixtral 8x7b                & 54.0\% & 65.3\% & \ding{52} & 38 &  2 & 12 & 31 &  6 & 13 & 21 & 25 & 21 & \$0.40 & \$269 \\
Qwen2.5 7B                  & 54.7\% & 65.7\% & \ding{52} & 47 & 26 & 3 & 36 & 24 & 8 & 20 & 26 & 22 & \$0.04 & \$27 \\
Gemma-3 12B                 & 55.7\% & 80.0\% & \ding{52} & 50 & 27 & 0 & 44 & 24 & 0 & 19 & 27 & 23 & \$0.04 & \$27 \\
Claude-3.5 Haiku            & 56.0\% & 79.0\% & \ding{55} & 50 & 12 & 0 & 44 & 42 & 0 & 18 & 28 & 24 & \$0.08 & \$54 \\
Mistral 7B                  & 59.0\% & 76.7\% & \ding{52} & 46 & 11 & 4 & 40 & 13 & 4 & 17 & 29 & 25 & \$0.03 & \$20 \\
Llama-4 Scout               & 59.0\% & 80.0\% & \ding{52} & 50 & 43 & 0 & 44 & 20 & 0 & 16 & 30 & 26 & \$0.08 & \$54 \\
Ministral 3B                & 49.3\% & 56.7\% & \ding{52} & 37 &  2 & 13 & 23 & 11 & 21 & 14 & 32 & 28 & \$0.04 & \$27 \\
\bottomrule
\end{tabular}
\smallskip

\footnotesize
Legend: \textbf{EMA} = Exact-match accuracy; \textbf{ONE} = At least one correct.  
\textbf{O} = Open-source; \textbf{MT} = Cost per million input tokens; \\
\textbf{ALL} = Cost estimated (cleaned dataset with input+output tokens) if the given model would be used for all 100,000 articles;  \\
Cost estimation is based on \url{https://openrouter.ai/models} pricing as of October 2025. 
\label{tab:bench_results}
\end{table*}

The benchmarking results in Table~\ref{tab:bench_results} show that GPT-5 achieved the highest exact-match accuracy (EMA) at 82.7\%. Among open-weight models, Qwen-2.5 72B and Llama-3.1 405B performed strongest at 80.0\% EMA, reducing the gap to proprietary systems to less than three percentage points. On the ONE metric (at least one correct extraction), Gemini 2.5 Pro performed best at 93.3\%, while Qwen-2.5 72B reached 91.0\%, further demonstrating the competitiveness of open-weight alternatives.

Most  models show strong performance on CRED. In contrast, NETID artifacts exhibit substantially higher false-positive and false-negative rates, suggesting that contextual infrastructure identifiers require deeper semantic discrimination. PII detection lies between these extremes and is primarily constrained by over-extraction rather than omission.
The gap between EMA and ONE indicates that many models detect at least one sensitive artifact per sample but struggle with complete multi-entity extraction. 

A notable finding is that TruffleHog, unlike the evaluated LLMs, detected only four true credentials while producing numerous false positives, resulting in EMA and ONE scores below 10\%. This underscores the limitations of purely pattern-based secret scanners for context-sensitive artifact extraction in archival source data.

Cost becomes decisive when extrapolating to the full corpus. Based on OpenRouter pricing (October 2025), GPT-5 input tokens cost \$1.25 per million, whereas Qwen-2.5 72B costs \$0.07 per million (approximately 96\% lower). With approximately 224 million input tokens across cleaned comments from 100{,}000 papers, this corresponds to an estimated \$15.68 for Qwen-2.5 72B versus over \$280 for GPT-5 for input tokens alone. Including output tokens and non-cached system prompts increases total cost substantially. Notably, the 2.7 percentage point EMA advantage of GPT-5 over Qwen-2.5 72B corresponds to an order-of-magnitude increase in inference cost, indicating diminishing returns at the upper end of the performance spectrum.

Based on this combined performance–cost analysis, we selected Qwen-2.5 72B for large-scale processing of arXiv submissions. Using parallel execution via the OpenRouter API, the model processes the 300-entry \texttt{LLMSec-DB} benchmark in 15--20 seconds. Full analysis of the LaTeX comment corpus requires over five hours and costs approximately \$50.4, including system prompts and output tokens. All model outputs were stored as structured JSON for downstream analysis and targeted manual validation. Illustrative cases are presented in Section~\ref{sec:discussion} while ensuring that individual authors remain undisclosed. The exact \href{https://github.com/LaTeXpOsEd/LaTeXpOsEd/blob/main/resources/system-prompt.md}{system prompt} and \href{https://github.com/LaTeXpOsEd/LaTeXpOsEd/blob/main/3_mine_entity-extraction.ipynb}{Jupyter notebook} used for entity extraction are publicly available in the project repository.

\subsubsection{Logical Filtering} \label{sec:logical}
As the third step, we identify non-LaTeX source files that may contain sensitive information via logical filtering. We reconstruct a dependency graph of each submission by mapping every file referenced during compilation of the final PDF. This yields the set of assets explicitly included in the published document. We exclude these referenced files and retain only files that are present in the source bundle but never imported into the final output. Such unreferenced files typically arise for three reasons:
\begin{enumerate}
    \item \textbf{Stale or superseded artifacts:} temporary placeholders, intermediate exports, or files replaced by newer versions;
    \item \textbf{Internal authoring aids:} private notes, draft fragments, or formatting guides used during writing;
    \item \textbf{Collaborative-platform byproducts:} artifacts exported from tools such as Overleaf (e.g., measurements, discussions, or procedural notes) that are not intended for publication.
\end{enumerate}
Unreferenced files constitute a distinct disclosure surface because they are not visible in the compiled PDF but remain publicly accessible within the source bundle.

We filter by file type to reduce noise and focus on formats that plausibly contain secrets. We remove file classes unlikely to expose sensitive data, including figure and chart sources (e.g., \texttt{.eps}), bibliographic files (\texttt{.bib}, \texttt{.bbl}), and styling/rendering artifacts (e.g., \texttt{.pygtex}), as well as similar build byproducts. After this step, the dataset is reduced to approximately 70{,}000 candidate files. These include structured data files (\texttt{.xml}, \texttt{.json}, \texttt{.csv}, \texttt{.html}), plain-text documents (\texttt{.txt}, \texttt{.md}, \texttt{.list}), source code (\texttt{.py}, \texttt{.js}), and logs/backups (\texttt{.log}, \texttt{.bkp}, \texttt{.db}), among others.

At this stage, we reapply techniques from Phase~1. Specifically, we run TruffleHog together with 1{,}750 regular expressions from the Secret Patterns Database, and we extract metadata (EXIF) where applicable to recover timestamps, authoring information, editing history, and potential location traces. All detected matches are subsequently subjected to manual validation to eliminate false positives.

\subsection{Phase 4 - Analysis}

In the final phase, we applied automated post-processing filters to reduce false positives. All extracted findings were stored as \texttt{.jsonl} records and analyzed with Python scripts to produce aggregate statistics and the charts/tables reported in this paper. We prioritized manual review for high-impact categories, in particular  login credentials and exposures involving substantial amounts of PII.

For critical cases including private authentication data, we attempted to identify contact information and directly notified the affected authors. For a wide category of findings, effective notification methods remain limited. To mitigate privacy risks and prevent misuse, we anonymized and sanitized all outputs to ensure that results cannot be reverse-engineered to recover the exact set of scanned archives. We discuss representative cases and their implications in Section~\ref{sec:discussion}.

\section{Discussion \& Results}
\label{sec:discussion}

In this section, we present the key findings from our large-scale analysis of $100{,}000$ arXiv submissions. We utilize both traditional tools and techniques such as TruffleHog and regular expressions to detect secrets in comments and files, along with advanced LLMs.

\subsection{Traditional Methods of Secret Detection}
\begin{figure}[b]
  \centering\includegraphics[width=1\columnwidth,page=1]{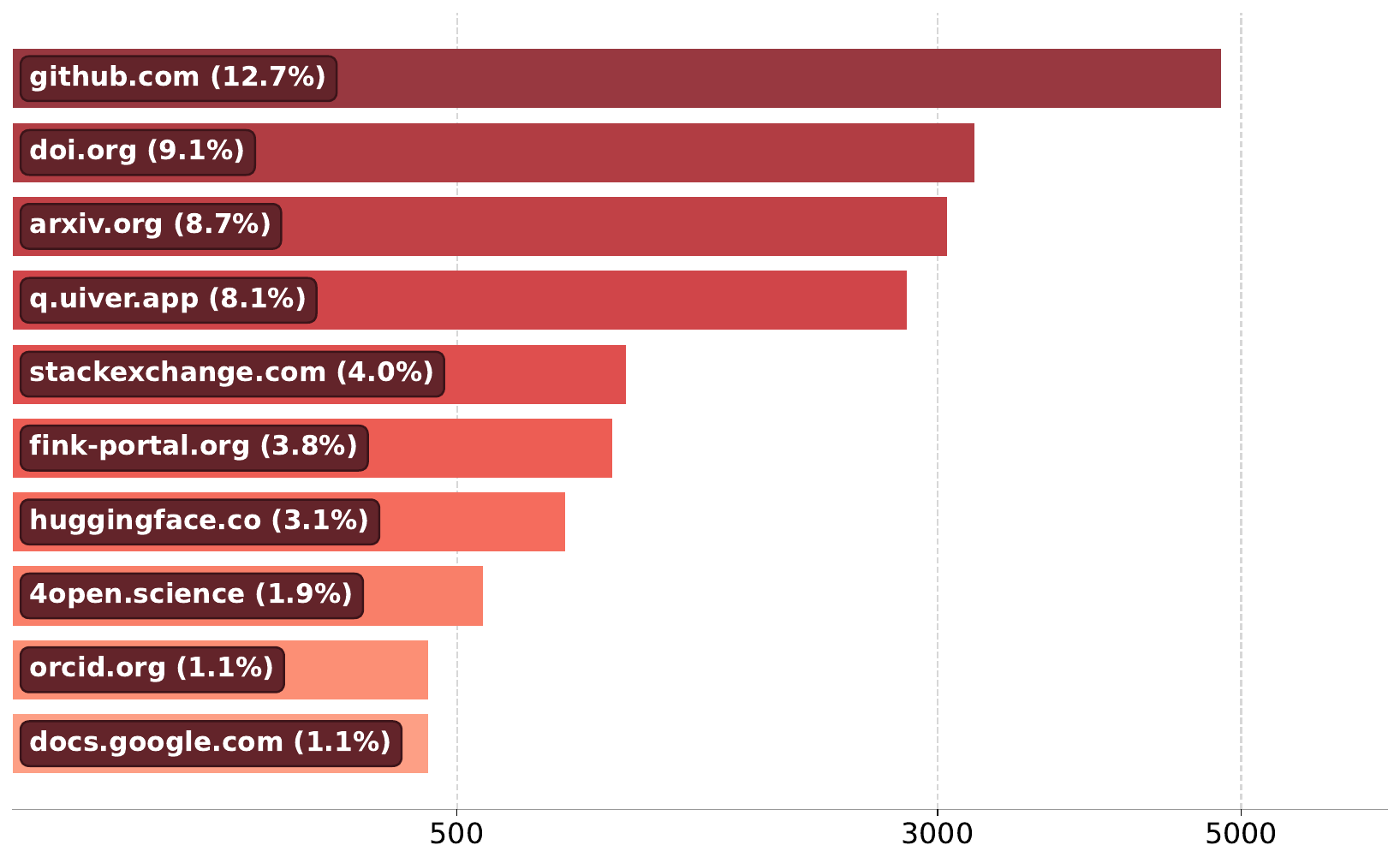}
  \caption{The ten most commonly occurring domains in URLs extracted from the comments.}
  \label{fig:common_domains}
\end{figure}

Using simple pattern-matching techniques and regex searches, we extracted approximately 42,500 unique URLs, including IP addresses, from the LaTeX comments. We performed ISP and domain reverse lookups on all collected addresses and found that 22\% originated from the United States, 9\% from France, and 7\% from China, with the top ISPs being Orange, Level3, and Amazon. The most common domains are presented in Figure~\ref{fig:common_domains}.

Although these domains are noteworthy, they rarely contain sensitive information. Instead, such data more often appears on file-sharing portals or private websites, sometimes with access tokens embedded directly in the URL. We identified 206 IP addresses---only eight in private ranges---with ports suggesting web servers, databases, FTP (20/21), SSH (22), unencrypted web services (80, 8080, 8000, 4321), printers (9100), and other services. Note, that ports are inferred from the string format, not from probing. Even more concerning, we discovered a private online chat platform in which login credentials were embedded in the URL, effectively making the entire chat history publicly accessible.

We also identified over 700 links with long token-like paths or query parameters, indicating that many may grant access without additional authentication. Using simple pattern matching, we collected more than 30 CASA tokens granting access to scientific publications; these are typically intended for internal sharing within institutions and not for public exposure. Furthermore, we found nearly 650 links granting view or edit access to private files and folders on common cloud storage services (e.g., Google Drive, Dropbox, Proton Drive, Mega). Approximately two thirds of the linked content was accessible without further authentication. 
Leaked documents fall into the following categories:
\begin{enumerate}[label=(\roman*), leftmargin=*, itemsep=0.25em]
  \item \textbf{Peer-review material:} reviews, rebuttals, and discussions.
  \item \textbf{Internal correspondence:} private exchanges about results or methodology.
  \item \textbf{Internal systems:} links to repositories, development environments, or shared dashboards.
  \item \textbf{Draft research artifacts:} unpublished manuscripts and files with embedded comments.
  \item \textbf{Data/model artifacts:} unreleased datasets and model weights.
  \item \textbf{Signed-link chains:} documents containing additional URLs enabling lateral access.
  \item \textbf{Spreadsheets and records:} experimental measurements and survey data (sometimes including PII).
  \item \textbf{Other private material:} letters, plans, internal documentation, and supplementary media.
\end{enumerate}

It is also important to highlight that most of the shared documents also allow editing, as they were likely shared for easy collaboration.

\subsection{LLM-Based Methods for Secret Detection}
The LLM-based analysis produced substantially more nuanced and higher-severity findings than the rule-based baseline. As described in the methodology section, we employed Qwen-2.5 72B to identify hidden secrets, leaked credentials, and internal author discussions that are unlikely to be detected through simple pattern matching or regular expressions alone. Many of these disclosures are implicit, fragmented across files, or only recognizable when interpreted in context (e.g., surrounding explanations, access instructions, or conversational cues), a task for which current LLMs are well-suited.

Across the dataset, Qwen-2.5 72B flagged $9{,}926$ papers. Because individual papers may contain multiple issues, this resulted in a total of $13{,}806$ detections. While PII, conflict, and peer-review related findings dominate the overall distribution, the model also identified $47$ instances of explicit authentication secrets (i.e., passwords), along with other access-enabling artifacts.
\begin{table}[t]
\footnotesize
\centering
\caption{Sanitized examples of credential leaks in arXiv submissions.}
\rowcolors{2}{gray!20}{white}
\renewcommand{\arraystretch}{1.25}
\setlength{\tabcolsep}{6pt}

\begin{tabular}{p{0.3cm}|p{2.5cm}|p{4.5cm}}
\rowcolor{gray!60}
\multicolumn{1}{c|}{\textbf{ID}} & 
\multicolumn{1}{c|}{\textbf{Type}} & 
\multicolumn{1}{c}{\textbf{Sanitized Example}} \\
\toprule
\rowcolor{orange!40}
1 & Portal Login & \texttt{https://****.org}, Paper ID: \texttt{****}, Password: \texttt{[{*****}]} \\
\rowcolor{lightred!80}
\hline
2 & Email + password & Email: 
\texttt{w*****@p****.org.cn}, Password: \texttt{*****} \\
\rowcolor{lightred!80}
\hline
3 & Project website & \texttt{https://sso.****.com}, Email: \texttt{****@****.com}, PW: \texttt{[{*****}]} \\
\rowcolor{orange!40}
\hline
4 & Priv. SharePoint link & \texttt{https://u*****-my.sharepoint
.com/:u:/g/personal/[****]} \\
\hline
\rowcolor{orange!40}
5 & Web Login & Login: \texttt{https://****/login}, Username: \texttt{re***}, PW: \texttt{[*****]} \\
\hline
\rowcolor{lightred!80}
6 & Github account & **** \texttt{github.com/******} pass: [{[*****]}]\\
\hline
\rowcolor{lightred!80}
7 & Gmail account & **** \texttt{l*****gmail.com} pass: [{[*****]}]\\
\hline
\rowcolor{orange!40}
8 & Paper submission& Your paper number is: ***4. Your paper access password is: C********\\
\hline
\rowcolor{orange!40}
9 & Private docs& \texttt{https://docs.google.com/doc/} \texttt{/d/1jC2****ULw/edit} \\
\hline
\rowcolor{lightred!80}
10 & iCloud login& icloud un: \texttt{a****@***}, pw: \texttt{[***]} \\
\bottomrule
\end{tabular}

\smallskip
\textcolor{orange}{Orange} = moderately sensitive data; \textcolor{lightred}{Red} = critically sensitive data.
\label{tab:leaks}
\end{table}
Table~\ref{tab:leaks} presents sanitized examples of real-world credential leaks extracted from LaTeX comments. We contacted the authors, where possible, to warn them of a potential data privacy breach.

Table~\ref{tab:category_distribution} summarizes the distribution of detected label categories introduced in Section~\ref{sec:list}, excluding the \texttt{OTHER} category. PII-related findings dominate the distribution, followed by peer-review and conflict-related disclosures. Network identifiers and explicit authentication secrets occur less frequently but remain security-relevant.
The \texttt{CRED} category refers strictly to explicit authentication secrets (e.g., passwords). Tokens embedded in URLs, including those granting access to shared private folders, are categorized separately under sensitive URLs and identified via URL pattern matching.

To ensure reliability, we manually validated all detected credentials and network identifiers. In addition, we reviewed approximately 10\% of conflict-, peer-review-, and PII-related findings. Validation consisted solely of confirming true positives; we did not attempt to authenticate against any system or use any discovered credentials.
In several cases, exposed credentials were accompanied by the corresponding login URL, directly indicating the target system. This demonstrates that low-cost automated analysis (like under \$50 in LLM inference costs) can uncover actionable OSINT at scale.

\begin{table}[t]
\centering
\caption{Distribution of Label Categories Identified by Qwen-2.5 72B}
\rowcolors{3}{gray!20}{white}
\begin{tabular}{l|r}
\toprule
\rowcolor{gray!50}
\textbf{Category} & \textbf{Count} \\
\midrule
PII & 5689 \\
CONF, PEER & 2228 \\
PEER, PII & 518 \\
CONF & 383 \\
CONF, PII & 331 \\
CONF, PEER, PII & 330 \\
PEER & 294 \\
NETID, PII & 71 \\
CRED, PII & 37 \\
NETID & 17 \\
CRED & 10 \\
CRED, NETID, PEER & 5 \\
NETID, PEER & 4 \\
CRED, NETID, PEER, PII & 4 \\
NETID, PEER, PII & 3 \\
CRED, NETID & 1 \\
CRED, NETID, PII & 1 \\
\midrule
\rowcolor{gray!50}
\multicolumn{2}{c}{\textbf{Overall Distribution by Category (summarized)}} \\
\midrule
PII & 6984 \\
PEER & 3386 \\
CONF & 3283 \\
NETID & 106 \\
CRED & 47 \\
\midrule
\textbf{SUM} & \textbf{13806} \\
\bottomrule
\end{tabular}
\label{tab:category_distribution}
\end{table}

\subsection{Logical Filtering of Unreferenced Files}

In multiple cases, credentials that had been removed from the visible manuscript persisted in auxiliary artifacts such as logs, backup files, or intermediate build outputs included in the submission archive. These residual disclosures were not present in the compiled PDF but remained accessible through the source bundle. Structured data formats proved particularly revealing. Configuration files (e.g., JSON or YAML) exposed service endpoints, authentication parameters, and internal infrastructure identifiers. Database artifacts and source code files contained embedded tokens, debugging traces, and system-level references. Unlike comment-based disclosures---which often reflect informal author communication---these artifacts represent workflow residue generated by collaborative tooling, version control, or automated build processes.

Metadata extraction introduced an additional layer of exposure. Document properties and embedded EXIF data revealed authoring timestamps, local usernames, software environments, and, in some cases, precise geolocation coordinates---information that can meaningfully compromise privacy or operational security~\cite{10.5555/1924931.1924933}.

\begin{table*}[t]
\footnotesize
\centering
\caption{Summary of Sensitive Information Disclosures with Precise Regulatory Mapping}
\rowcolors{3}{gray!20}{white}
 \renewcommand{\arraystretch}{1.4} 
\begin{tabular}{l|l|l|r|l|l}

\toprule
\rowcolor{gray!50}
\textbf{Description} & \textbf{Source} & \textbf{Detection} & \textbf{Count*} & \textbf{Severity} & \textbf{Primary References} \\
\bottomrule
Login credentials (username/password) & comments & EE & 47 & \cellcolor[HTML]{6A0DAD}\textcolor{white}{Critical} & CWE-522, CWE-798, NIST SP 800-63B  \\
Large-scale (aggregated) PII exposure & links & PM & 22 & \cellcolor[HTML]{6A0DAD}\textcolor{white}{Critical} & FIPS 199, NIST SP 800-122 \\
Survey data with large-scale PII exposure & links & PM & 2 & \cellcolor[HTML]{6A0DAD}\textcolor{white}{Critical} & FIPS 199, GDPR Art. 32(1), NIST SP 800-122 \\
AWS access keys & comments & PM & 2 & \cellcolor[HTML]{6A0DAD}\textcolor{white}{Critical} & CWE-522, CWE-798,  NIST SP 800-53 AC-3 \\
JSON Web Tokens (JWTs) & comments & PM & 5 & \cellcolor[HTML]{6A0DAD}\textcolor{white}{Critical}  & CWE-613, OWASP Top 10 A07\\
\hline
Social security numbers & comments & PM & 26 & \cellcolor[HTML]{CC0119}\textcolor{white}{High} & NIST SP 800-122, FTC Identity Theft guidance \\
PII Exposure & comments & EE & 6984 & \cellcolor[HTML]{CC0119}\textcolor{white}{High} & NIST SP 800-122, GDPR Art. 4 + Recital 75 \\

Image location EXIF data & files & LF & 631 & \cellcolor[HTML]{CC0119}\textcolor{white}{High} & GDPR Recital 30, 5th USENIX Conference~\cite{10.5555/1924931.1924933}\\
Private documents and folders & links & PM & 129 & \cellcolor[HTML]{CC0119}\textcolor{white}{High} & ISO 27002:2022 5.12\\
Private source code & links, files & PM, LF & 2774 & \cellcolor[HTML]{CC0119}\textcolor{white}{High} & ISO 27002:2022 8.28\\

\hline
Peer review disputes / disagreements & comments & EE & 3386 & \cellcolor[HTML]{F29026}\textcolor{black}{Medium} & ISO 27002:2022 5.12 \\
Author conflicts / false results & comments & EE & 3283 & \cellcolor[HTML]{F29026}\textcolor{black}{Medium} & GDPR Recital 75, COPE guidelines \\
Private Git repositories (HTTP/SSH) & comments & PM & 658 & \cellcolor[HTML]{F29026}\textcolor{black}{Medium} & ISO 27002 8.28 \\
Phone numbers (non-public) & comments & PM & 60 & \cellcolor[HTML]{F29026}\textcolor{black}{Medium} & NIST 800-122, GDPR Recital 75  \\
CASA tokens (publication access) & comments & PM & 30 & \cellcolor[HTML]{F29026}\textcolor{black}{Medium} & CWE-522\\
Private model weights / training data & links, files & PM, LF & 28 & \cellcolor[HTML]{F29026}\textcolor{black}{Medium} & ISO 27002:2022 8.28 \\

SQLite databases & files & LF & 7 & \cellcolor[HTML]{F29026}\textcolor{black}{Medium} & ISO 27002:2022 8.10 \\
\hline
Unintended IP address exposure & comments & PM & 198 & \cellcolor[HTML]{FFD51C}\textcolor{black}{Low} & NIST SP 800-115 \\
Unique email addresses & comments & PM & 8236 & \cellcolor[HTML]{FFD51C}\textcolor{black}{Low} & NIST 800-122 \\
Structured config files (JSON/YAML) & files & LF & 802 & \cellcolor[HTML]{FFD51C}\textcolor{black}{Low} & NIST SP 800-53 CM-2 \\
Hashes (MD5, SHA1, SHA256) & files, comments & LF, PM & 549 & \cellcolor[HTML]{FFD51C}\textcolor{black}{Low} & CWE-212 \\
Network \& Device IDs (MAC, Host ID) & comments & EE & 106 & \cellcolor[HTML]{FFD51C}\textcolor{black}{Low} & NIST SP 800-115 3.3.1, GDPR Recital 30 \\
Validated IBANs & comments & PM & 92 & \cellcolor[HTML]{FFD51C}\textcolor{black}{Low} & GDPR Recital 75, NIST 800-122 \\

P.O Box addresses & comments & PM & 75 & \cellcolor[HTML]{FFD51C}\textcolor{black}{Low} & NIST SP 800-122 Sec. 4.2.2 \\
\bottomrule
\end{tabular}

\smallskip

\footnotesize
\textbf{EE}: Entity Extraction, \textbf{PM}:Patter Matching, \textbf{LF}: Logical Filtering

\label{tab:total_findings}
\end{table*}

\subsection{Summary of Findings}

The relevance of these findings is grounded in the principle of informational self-determination~\cite{lazaro2015control} and Westin’s definition of privacy as ``\textit{the ability of individuals or institutions to determine when, how, and to what extent information about them is communicated to others}''~\cite{westin1967privacy}. Although arXiv’s submission agreement grants an ``\textit{irrevocable}'' and ``\textit{perpetual}'' license to distribute submitted materials and places responsibility for sanitization on authors~\cite{arxiv_submission_agreement, arxiv_terms, arxiv_privacy}, formal consent to distribution does not necessarily amount to awareness and meaningful informational control.

\subsubsection{Severity Evaluation}

Table~\ref{tab:total_findings} provides an artifact-level classification of disclosures. Rather than grouping findings solely by semantic labels (e.g., PII, CONF, PEER), the table enumerates concrete exposure types (e.g., login credentials, AWS keys, EXIF metadata, SQLite databases). Severity levels reflect a composite, risk-based assessment that integrates confidentiality impact and practical exploitability. This assessment draws on regulatory frameworks, technical security standards, vulnerability taxonomies (e.g., CWE), peer-reviewed literature, and industry best-practice guidance (e.g., OWASP). While CWE-200 (Exposure of Sensitive Information) is broadly applicable across many disclosure types, it is not redundantly cited for each row in order to preserve specificity of reference mapping.

\textbf{Critical findings} enable immediate unauthorized access, credential misuse, or large-scale compromise without requiring an additional vulnerability. These include explicit login credentials, AWS access keys, authentication tokens, and large-scale PII exposures accessible through public links. Such artifacts represent authentication and access-control failures and may trigger breach-notification obligations under GDPR Article~34. 

\textbf{High-severity findings} substantially increase the likelihood of identity misuse, targeted social engineering, intellectual property loss, or meaningful privacy harm. These include general PII exposure, private document repositories, image metadata containing embedded GPS coordinates, and other artifacts that elevate exposure risk without directly granting system access.

\textbf{Medium-severity findings} concern artifacts that primarily facilitate reconnaissance, targeted follow-on exploitation, reputational damage, or misuse within a narrower threat model. These include peer-review disputes, author conflicts, private Git repositories, non-public phone numbers, CASA tokens, private model weights or training data, and embedded SQLite databases.

\textbf{Low-severity findings} consist of contextual identifiers and technical artifacts whose primary risk emerges through aggregation, correlation, or secondary misuse. These include unintended IP address exposure, IBANs, unique email addresses, structured configuration files without embedded secrets, cryptographic hashes and P.O. Box addresses. 

\subsubsection{When Public Data Becomes a Privacy Risk}
A natural question arises as to why data elements such as IBANs and P.O. Box addresses are classified as low severity, given that they are often intended for public or administrative use and therefore regarded as non-sensitive. Yet privacy risk cannot be assessed solely on that assumption; what matters is the context of disclosure and whether publication was intentional.

When such information appears in commented-out sections of LaTeX files or in auxiliary source materials not visible in the compiled PDF, its placement indicates that the authors did not intend it for public communication. If these materials are nevertheless archived and made accessible, this results in unintended disclosure and a loss of control over personal data. Moreover, when combined with names, email addresses, or institutional affiliations, such data can facilitate profiling, targeted attacks, or social engineering. By contrast, when authors deliberately include this information in the visible document for public-facing purposes, the disclosure is intentional and does not raise the same privacy concerns.

This dynamic reveals a structural gap between formal authorization and practical awareness, as disclosures occur in materials invisible to standard reading workflows despite authors’ agreement to broad distribution terms.

\section{Limitations and Threats to Validity}
\label{sec:limit}
We have identified the following limitations of our study:
\begin{itemize}

 \item  \textbf{Scope of submission formats.} \texttt{LaTeXpOsEd} targets arXiv source bundles that contain LaTeX (and associated auxiliary assets such as images, bibliography files, and build artifacts). Consequently, disclosures that occur exclusively in non-LaTeX workflows (e.g., Word-based manuscripts, or pipelines that only publish a flattened PDF without source) are out of scope. 

 \item  \textbf{Detection and extraction error modes.} Our pipeline combines lightweight heuristics (e.g., URL harvesting and structural filters) with LLM-based entity extraction. Both components introduce errors: heuristics can miss non-standard patterns, while LLMs can produce false positives on contextually benign strings (e.g., test credentials or placeholders) and false negatives when secrets are fragmented, obfuscated, or presented in atypical formats. In particular, encoded representations (e.g., Base64, hexadecimal, compressed blobs) and multi-step reconstructions are not handled reliably, which likely leads to undercounting of certain exposure classes.

 \item  \textbf{Sampling and corpus generalization.}
    Our dataset comprises 100{,}000 submissions (\mbox{\textasciitilde}3.6\% of arXiv at the time of measurement) drawn via uniform random sampling, which reduces the risk of systematic bias across time periods and subject areas. Nevertheless, it is not a full census, and corpus-wide extrapolations remain sensitive to residual sources of uncertainty (e.g., temporal drift in authoring practices, heterogeneous comment usage, and the fact that only the subset of papers with usable comments can be analyzed). Importantly, our controlled benchmark and targeted manual validation confirm that the exposures we report are genuine and that the phenomenon is not an artifact of our pipeline, even if exact population-level rates may vary under full-corpus processing.

 \item    \textbf{Cost and scalability:} On our 100{,}000-paper dataset, the variable cost of the pipeline was approximately \$37 for arXiv S3 retrieval plus \$50.4 for Qwen-based comment analysis, for a total of \mbox{\textasciitilde}\$90 per 100{,}000 papers. Extrapolating the same S3+Qwen pipeline to the full arXiv corpus, and noting that only 81.6\% of papers contain usable comments after preprocessing, yields an expected cost of roughly \mbox{\textasciitilde}\$2.2k for comment analysis at scale \mbox{($0.816 \times 30 \times \$50.4$)} plus \mbox{\textasciitilde}\$1.1k for full-corpus S3 retrieval \mbox{($30 \times \$37$)}, i.e., \mbox{\textasciitilde}\$3.3k total. As a point of comparison, using GPT-5-mini instead of Qwen increases the LLM component: based on our token measurements, input-side costs for 100{,}000 papers are approximately \$340, and including outputs and non-cached prompts can increase this by roughly a factor of three, yielding \mbox{\textasciitilde}\$1{,}020 per 100{,}000 papers. Under the same assumption that only the usable-comment subset is processed by the model, the full-arXiv LLM cost would be approximately \mbox{\textasciitilde}\$25k \mbox{($0.816 \times 30 \times \$1{,}020$)}, plus \mbox{\textasciitilde}\$1.1k for S3 retrieval, i.e., \mbox{\textasciitilde}\$26k total.

\end{itemize}

\section{Conclusion}
\label{sec:conclusion}

This paper introduced \texttt{LaTeXpOsEd}, a framework that combines traditional artifact extraction with modern language-model reasoning to detect and characterize sensitive information leakage in public preprint source packages. Evaluating a representative sample of 100{,}000 arXiv submissions, we find that inadvertent disclosure is not an edge case: our pipeline flagged 9{,}926 papers (9.9\%), producing 13{,}806 distinct detections spanning privacy, security, and research-integrity risks.

Beyond prevalence, the results reveal concrete, recurring exposure modes. We observe large volumes of personally identifying information (e.g., names, emails, and phone numbers) and substantial leakage of peer-review and editorial materials, alongside internal author communications that could lead to organisational or individual reputation damage. We also identify link-based access leakage (e.g., private documents/folders and shared URLs) and a smaller but consequential set of explicitly \emph{critical} cases, including 47 username/password credentials and cloud access keys. Taken together, these findings show that public source release can unintentionally turn build artifacts, comments, and auxiliary files into high-value OSINT.

Importantly, we conducted a systematic benchmark of 25 LLMs for context-sensitive leak detection, and selected Qwen as the best-performing open-source model for \texttt{LaTeXpOsEd} based on the resulting accuracy-cost trade-offs. To support reproducibility and follow-on research, we also publicly released our benchmark (models, prompts, and evaluation protocol), providing a concrete reference point for future work on LLM-assisted sensitive-data detection in scientific source archives.

\noindent\textbf{Recommendations.}
Mitigation requires both author-side hygiene and platform-side safeguards.
For authors (pre-submission hygiene) we recommend the following action points:
\newpage
\begin{enumerate}
    \item \textbf{Sanitize the source package.} Remove comments, unused inputs, and non-essential artifacts before upload; where possible, use automated tooling such as \textit{arXiv LaTeX Cleaner}\footnote{\url{https://github.com/google-research/arxiv-latex-cleaner}}.
    \item \textbf{Keep secrets out of the project tree.} Never store credentials, tokens, private links, or ``notes-to-self'' in \texttt{.tex} comments, build logs, auxiliary files, or temporary caches.
    \item \textbf{Avoid link-based access leakage.} Do not circulate editable signed URLs or shared links that implicitly grant access; share access explicitly with co-authors via proper access control.
    \item \textbf{Audit before submission.} Perform a final directory audit (including \texttt{.log}, \texttt{.aux}, \texttt{.bbl}, and embedded metadata such as image EXIF).
\end{enumerate}

\noindent\textbf{For platforms (systemic safeguards):}
\begin{enumerate}
    \item \textbf{Make irreversibility explicit.} Present clear upload-time warnings that all included source files become public and are difficult to retract once released.
    \item \textbf{Provide safe preprocessing.} Offer optional (or default) sanitization such as stripping comments and excluding unused files and common build artifacts.
    \item \textbf{Scan for sensitive data at upload time.} Run credential/secret/PII detectors and provide immediate, actionable feedback to authors before publication.
    \item \textbf{Revisit source-release defaults.} Where feasible, consider staged release or restricted access to source packages for elevated-risk submissions.
\end{enumerate}

Finally, our study addressed the two guiding research questions:
\begin{tcolorbox}[colback=gray!10, colframe=black, arc=6pt, boxrule=0.7pt,
  left=2mm, right=2mm, top=1mm, bottom=1mm, rounded corners]
\textbf{RQ1: What kinds of sensitive information are exposed in preprint source files and comments, and how often do these disclosures occur?}\\
\textbf{Answer:} Across a corpus of 100{,}000 preprints, 9.9\% contained at least one sensitive disclosure. The exposed artifacts span multiple severity levels. High and critical impact findings include explicit credentials, cloud access keys, authentication tokens, and large-scale PII exposures, Social Security numbers, private documents, source code, and image metadata revealing geolocation. Medium and low-severity artifacts expose peer-review disputes, internal repositories, validated financial identifiers, network identifiers, and unique email addresses. These results indicate that sensitive disclosures are not isolated anomalies but occur across a broad spectrum of artifact types, affecting both individual privacy and institutional reputation
.
\end{tcolorbox}

\begin{tcolorbox}[colback=gray!10, colframe=black, arc=6pt, boxrule=0.7pt,
  left=2mm, right=2mm, top=1mm, bottom=1mm, rounded corners]
\textbf{RQ2: How effective are traditional pattern-matching techniques and LLMs at detecting contextually hidden sensitive data within preprint source materials?}\\
\textbf{Answer:} Pattern matching is effective for structured indicators (e.g., URLs) but can produce many false positives for ``secret-like'' strings. In contrast, LLM-based contextual reasoning improves precision in validation by using surrounding context to distinguish true disclosures from benign text, at higher computational cost.
\end{tcolorbox}

Overall, these results motivate urgent community awareness, and platform-level interventions so that the benefits of open dissemination are not undermined by preventable privacy and security failures. Our large-scale automated analysis captures systematic disclosure patterns but does not account for targeted, manual adversarial review. Such review may identify additional sensitive content within auxiliary artifacts beyond the detection scope of a single-pass automated pipeline.

\section*{Acknowledgement}
This research is supported and funded by ZEISS Digital Innovation and the Technology Innovation Institute (TII), Abu Dhabi. Additional support is provided by the TKP2021-NVA Funding Scheme under Project TKP2021-NVA-29; ELTE-OTP Cyberlab—a collaboration between Eötvös Loránd University (ELTE) and OTP Bank Plc; funding from Horizon Europe under Grant Agreement No. 101120853; the Research Council of Norway through Project No. 312122, ``Raksha: 5G Security for Critical Communications.'';and funding under Grant Agreement No. 101145874, supported by the European Cybersecurity Competence Centre.

\section*{Availability}

The LaTeXpOsEd project is publicly available on its GitHub project page: \url{https://github.com/LaTeXpOsEd}. Additionally, both the tools and the LLM-SecDB benchmark are archived on Zenodo at the following links:
\begin{itemize}
\item LLM-SecDB benchmark: \url{https://doi.org/10.5281/zenodo.20035860}
\item Arxiv secret detection tool: \url{https://doi.org/10.5281/zenodo.20058989}
\end{itemize}

\bibliographystyle{plain}
\balance
\bibliography{references}

\end{document}